\documentstyle[aps,twocolumn,epsf]{revtex}
\def\ave#1{\langle #1\rangle}

\newcommand{\op}[1]{\hat{#1}}

\newcommand{\bra}[1]{\langle #1|}
\newcommand{\ket}[1]{|#1\rangle}
\newcommand{\braket}[2]{\langle #1|#2\rangle}
\newcommand{\half}{\textstyle{\frac{1}{2}}}
\newcommand{\sgn}{{\rm sgn}}
\begin{document}
\title{THE QUANTUM MECHANICS OF CHAOTIC BILLIARDS}
\author{Giulio Casati}
\address{International Center for the Study of Dynamical Systems,
via Lucini,3, I--22100 Como, ITALY,\\ 
Istituto Nazionale di Fisica della Materia and INFN, Unit\`a di
Milano}
\author{Toma\v z Prosen}
\address{Physics Department, Faculty of Mathematics and Physics,
University of Ljubljana, Jadranska 19, 1111 Ljubljana, Slovenia
}
\date{\today}
\draft
\maketitle
\begin{abstract}
We study the quantum behaviour of chaotic billiards which
exhibit classically diffusive behaviour. In particular we consider
the stadium billiard and discuss how
the interplay between quantum localization  and the rich 
structure of the classical phase space influences the 
quantum dynamics. The analysis of this model leads to new 
insight in the understanding of quantum properties
of classically chaotic systems.
\end{abstract}

\section{INTRODUCTION}

When in a cold winter morning at the start of 1976 one of us (GC) entered
the Institute of Nuclear Physics of the Siberian Division of the USSR
Academy of Sciences in Akademgorodok, the main purpose was to discuss with
the author Boris V. Chirikov, a startling paper which appeared as a CERN
preprint at the end of the sixties\cite{CERN}. Except to few, the paper
remained
almost unnoticed and practically none had the perception that this paper
was a cornerstone of the new building of nonlinear science which, only
 few years later, expanded in an impressive way.
Among other things, the so-called standard map or Chirikov map was
discussed as a model of classical dynamical chaos. It was in that
same occasion that we quantized the map thus obtaining the quantum kicked
rotor which became later a basic model for the study of quantum
chaos\cite{CCFI,CC}.

    The importance of the Chirikov map lies in the fact that it gives an
approximate description of a wide class of dynamical systems. As a matter
of fact it contains most of the complex features of dynamical motion. For
these reasons its properties are continuously being studied and it is used
in university courses to teach the emergence of random motion in
deterministic systems.

In the quantum world the standard map played a fundamental role in
bridging different fields of research: solid state physics, atomic
physics, Random Matrix Theory. In particular it allowed to discover the
phenomenon of quantum dynamical localization namely the fact that quantum
interference effects suppress the classical diffusive process and lead to
localization of the quantum excitation. This phenomenon had then been
observed in several laboratory experiments in atomic physics\cite{CCGS,ba,ko,Wa}
 and, more
recently, a physical microscopic realization of the kicked rotator had
been possible \cite{Raizen}. 

All the above experiments refer to systems under external perturbations.
It was however not clear how quantum localization could take place in
conservative systems. Indeed in this case an Eigenfucntion (EF) is always a
superposition of a practically a finite number of unperturbed EFs and
therefore it is not even clear what localization should mean here.

In Refs.\cite{CCGI93,CCGI96}, 
by using a well known model of quantum conservative systems,
the so-called WBRM model, it was shown that EFs can be localized inside
the energy shell. More precisely, an ergodicity parameter $\lambda$ was
introduced which acts as a scaling parameter. For $\lambda<1$ the EFs 
are localized inside the energy shell $\delta E$, while for 
$\lambda>1$ the EFs are extended or ergodic. The width $\delta E$ is 
the width ot the Local Density of States (LDOS) or strength function. 
Therefore for $\lambda>1$
an EF actually excites the maximum number of unperturbed states. As it 
is quite evident this problem calls in question the notion of quantum
ergodicity \cite{cas}.

It was then shown that quantum localization actually takes place 
in a more
physical model of conservative systems: the billiard in a stadium \cite{BCL}.
The
same effect was successively confirmed in a different shape, 
classically chaotic, billiard \cite{FS1,FR}.

The question remains open however, whether or not the mechanism of
localization is the same as in WBRM. In particular it is necessary to
understand how the structure of EF changes as one moves from the
perturbative regime to the region of complete ergodicity.

In this paper, on the basis of the results of \cite{CP} , we discuss in detail
the above problem. We show that contrary to WBRM, localization here
manifests via the sparsity of the structure of the EFs and we surmise
that
this should be a general feature of localization in conservative systems.
We also show that the presence of cantori in the classical motion plays an
important role in the quantum motion. The ``quantum cantori'' should be
observable in real laboratory experiments.

\section{THE CLASSICAL STADIUM  BILLIARD}

In this  paper we mainly consider a {\em stadium billiard}, 
that is the motion of a free
point particle of unit mass and velocity $\vec{v}$ (energy
$E=v^2/2$) bouncing elastically inside a stadium-shaped well: 
two semicircles of radius $1$ connected by two straight line 
segments of length $2\epsilon$. 
The classical motion, for arbitrary small $\epsilon$, is ergodic, 
mixing and exponentially unstable with Lyapunov exponent 
$\Lambda \sim \epsilon^{1/2}$.
It can be approximated (up to ${\cal O}(\epsilon)$)
with the discontinuous stadium-map \cite{BCL}
\begin{eqnarray}
L_{n+1} &=& L_n - 
2\epsilon\sin\theta_n\sgn(\cos\theta_n)\sqrt{1-L_n^2}, 
\label{eq:stadiummap}\\
\theta_{n+1} &=& \theta_n + \pi - 2\arcsin{L_{n+1}}
\pmod{2\pi}
\nonumber
\end{eqnarray}
where $L=l/\sqrt{2E}$ is the rescaled angular momentum, 
$l = \vec{r}\wedge\vec{v}$, and $\theta$ is the polar angle 
(identical to the arc-length for small $\epsilon$). 
Note that, due to  symmetry, 
the stadium map (\ref{eq:stadiummap}) has a period $\pi$ in the
angle $\theta$.
One should note also that the stadium map 
is topologically similar to the sawtooth map
\begin{eqnarray}
p_{n+1} &=& p_n + K (x-\half) \label{eq:sawtoothmap} \\
x_{n+1} &=& x_n + p_{n+1} \pmod{1}. \nonumber
\end{eqnarray}
The behaviour of the sawtooth map (2) has been studied analytically
and numerically in great detail \cite{DANA,DANA1}.
For the stadium billiard, due to the classical chaotic motion, 
the  angular momentum $l$ undergoes, for small $\epsilon$, a process 
of diffusive behaviour
\begin{equation} 
\langle (l_n - l_0)^2 \rangle \approx D(l_0) n
\end{equation}
with diffusion rate $D=D(l)$ which may depend on local 
value of angular momentum $l$. Rigorous results on 
 the sawtooth-map \cite{DANA,DANA1} have shown that the diffusion 
rate goes as $D\propto K^{5/2}$ if $K \ll 1$ or 
$D\propto K^2$ if $K \gg 1$.
The same dependence $D\propto\epsilon^{5/2}$ 
on the small control parameter $\epsilon$ 
has been numerically found in \cite{BCL} for the
billiard and for
the map (\ref{eq:stadiummap}). The factor $\sqrt{1-L_n^2}$  
in the local shift of angular momentum 
leads to a  dependence of  diffusion rate on the local 
value of angular momentum. As a result, it is easily seen that,
for the map (\ref{eq:stadiummap}): 
\begin{equation}
D(l) = \gamma \epsilon^{5/2} (2E - l^2),
\label{eq:diff}
\end{equation}
where the prefactor  $\gamma \approx 2$ has been numerically computed.
In  figure 1 we
show the  weak dependence of  $\gamma$ on the local angular
momentum $l$ for various values of the parameter $\epsilon$ in the
stadium billiard.
Note that eq. (\ref{eq:diff}) gives the diffusion rate in the 
discrete time $n$. The physical time $T$ between collisions 
depends on the local angular momentum,
namely $T = \sqrt{2E - l^2}/E$.
Since the stadium billiard is mixing, the classical dynamics
will lead to the microcanonical equilibrium distribution. This implies 
that the equilibrium distribution 
of angular momenta is uniform in the discrete time
$$\rho_{\rm discr}(l) = 1/\sqrt{8E}$$
(invariant measure of the 
approximate Poincar\' e map (\ref{eq:stadiummap}))
while, in the continuous time, it obeys
the semicircle distribution
\begin{equation}
\rho(l) = \frac{1}{\pi E}\sqrt{2E - l^2} \propto T(l)\rho_{\rm discr} 
\label{eq:mc}
\end{equation}
(invariant measure of the smooth dynamics in the 
full phase space).

The power $5/2$ in eq. (\ref{eq:diff}) (instead of $2$ as found for
smooth deformations of the circle \cite{FS1}), 
is due to the existence of {\em cantori} in the classical
billiard motion  which are also typical of
discontinuous maps like (\ref{eq:stadiummap},\ref{eq:sawtoothmap})
The cantori form strong obstacles to phase space transport and therefore
they reduce the diffusion rate (even if the diffusion remains normal).

It is important to evaluate  the flux ${\cal F}$ of the
phase space area which is transported through a cantorus in one
iteration of the map. For the sawtooth map there are no dominant
cantori in phase space and the flux ${\cal F}$ is 
independent of the winding number of the
resonance and is given by 
${\cal F} = K^2 /8\sqrt{D}$ where $D = K^2 + 4K$ \cite{DANA}.

Therefore, for  small values of the control parameter $K$ the flux goes  
as $K^{3/2}$. 
We can now use  this result to estimate the flux through cantori for the 
stadium map for small $\epsilon$. We obtain
\begin{equation}
{\cal F} \approx (2E)^{1/2}\epsilon^{3/2}
\label{eq:flux}
\end{equation}
which incorporates the correct energy scaling of 
phase space area.

The size of a cantorus ${\cal C}$, in the rescaled angular momentum 
variable $L$, that is $p_{\cal C} = \max_{\cal C} L - 
\min_{\cal C} L$,
averaged over all the resonances, can be estimated from the
exact results on sawtooth map \cite{sizeofcantori}. After  
averaging over all the resonances one finds that the average size is
proportional to the parameter $\epsilon$, 
\begin{equation}
\bar{p}= c\epsilon,
\label{eq:sizeofc}
\end{equation}
where $c$ is some numerical constant. 
From our numerical computations on the stadium, $c$ turns out
to be in the range $c=10$ for eps=0.05, and $c=15$ for eps =0.005.
The fact that $c$ slowly increases with decreasing eps is due to
existence of the cantorus along the separatrix of 2:1
resonance (around $L=0$) which has a larger size and scales as
$p(2,1)\approx\sqrt{\epsilon}$.
In figure 2a we illustrate the structure of cantori by showing
a phase space portrait of a typical orbit
($\epsilon=0.003$) around the largest 2:1 resonance.

\section{THE QUANTUM DYNAMICS}

In this section  we consider the structure of
quantum eigenfunctions for the stadium billiard $\Psi_n(\vec{r})$
which are solutions of the Helmholz equation
$$(\Delta \Psi_n + k_n^2 \Psi_n) = 0.$$
with consecutive eigenenergies $E_n = k_n^2/2$.
We consider only odd-odd
states with Dirichlet boundary conditions on the quarter-stadium.
It is interesting to consider the eigenstates $\Psi_n$ of the
stadium billiard expressed in terms of eigenstates of the
nearest integrable billiard, namely the circle billiard.
The eigenfunctions of the unit quarter-circle billiard
are $$\Phi_{s m}(\vec{r}) = J_{2s}(k^0_{s m} r)\sin(2s\phi),$$
where $k^0_{s m}$ are zeros of even-order Bessel functions. One 
may expand an eigenstate of the stadium billiard (for small
\footnote{If $\epsilon$ is not really small, in
order to be able to express the eigenfunctions of the stadium in
terms of eigenfunctions of the circle, one should consider
the smallest enscribing circle billiard with radius $1+\epsilon$.} $\epsilon$)
in terms of eigenstates of a  quarter-circle billiard
\begin{equation}
\Psi_n(\vec{r}) = \sum_{sm} c^n_{sm} \Phi_{sm}(\vec{r})
\label{eq:express}
\end{equation}
The probability of having a value $l=2s$ of angular momentum (only
even values of angular momentum are allowed due to symmetry) is
\begin{equation}
p_n(l=2s) = \sum_{m} |c^n_{sm}|^2. 
\label{eq:pn}
\end{equation}
If quantum eigenstates were ergodic one would expect to 
recover the classical microcanonical distribution (\ref{eq:mc}) of 
angular momentum namely one would expect the quantum distribution
\begin{equation} 
p_n(l=2s) \approx \rho(l) \propto \sqrt{2E_n - l^2}.
\label{eq:qe}
\end{equation}

However, since the classical motion is approximated by the map (1), then
in analogy with the quantization of the standard map, one would 
expect  the phenomenon of quantum localization to take place with 
localization length $\ell$  proportional to the classical 
diffusion rate 
\begin{equation}
\ell \sim D.
\label{eq:lD}
\end{equation}
The localization length $\ell$, which is defined more precisely below
(eq. (\ref{eq:ll}), section 6), 
measures the (average) number of excited angular 
momentum eigenfunctions, that is the average width of the
probability distribution $p_n(l)$.

In a bound conservative system like the stadium however,
 there are two main peculiarities which influence the process
of quantum localization.
First, the range of variation of angular momentum $l$ is finite,
$-\sqrt{2E}\le l \le \sqrt{2E}$. 
As discussed in \cite{BCL}, this implies the existence of
an upper bound for the localization length  which is defined by the
condition $\ell_{\max} = l_{\max} = \sqrt{2E}$. This 
leads, together with (\ref{eq:lD}), to the {\em delocalization or 
ergodicity border} $\epsilon^5 N \sim 1$, where $N \approx E/8 = k^2/16$ 
is the sequential quantum number in the stadium billiard.
Therefore, in order to observe `dynamical' localization, this 
border must be above the {\em perturbative border}.
The perturbative border is given simply by the condition
$\epsilon \sqrt{2E} \sim 1$, which states  that the
length of the straight line segment of the stadium should be at least
one de Broglie wavelength in order to be
visible to quantum mechanics $\epsilon \sim (2E)^{-1/2}$. The same
border can be obtained from the
map (\ref{eq:stadiummap}) via the condition that in one iteration 
the change in angular momentum 
is at least one quantum $\Delta l \sim \epsilon \sqrt{2E} \sim 1$.
Therefore,  for small values of the parameter $\epsilon$ the two borders are
well separated 
$$ N_p = \frac{1}{16}\epsilon^{-2} \ll N_e = \frac{1}{64}\epsilon^{-5},
$$
and is possible to observe localization as actually done 
in \cite{BCL}.

The second peculiarity is due to the presence of cantori. Already long ago 
it has been surmised by 
McKay and Meiss \cite{MCKAY} that  cantori may act
as perfect barriers for the quantum motion if the flux through
cantori, for each iteration of the quantum map, is less than one Planck's cell. 
However, to our knowledge, this effect
 has never been observed
neither numerically nor experimentally and therefore it is not really known
whether the above condition actually plays a role in quantum mechanics.
For our present case of the stadium map this border can be estimated from the 
flux (\ref{eq:flux}) 
${\cal F} \sim x:=\epsilon^{3/2}\sqrt{2E} \sim 1$ 
(note that we have $\hbar = 1$), and gives  
\begin{equation}
N_c = \frac{1}{16}\epsilon^{-3}.
\end{equation}
It is interesting to observe that, for sufficiently small $\epsilon$, 
this border is well separated
from the other two borders $N_p \ll N_c \ll N_e$ and should be
(numerically or experimentally) observable.
Hence, for small values of the control parameter $\epsilon$, we expect to
see four different regimes in the quantum dynamics of the stadium biliard: 
(1) perturbative regime for $N < N_p$, 
(2) pseudo-integrable or cantori regime  where eigenstates are expected to
be localized on classical cantori, for $N_p < N < N_c$,
(3) dynamical localization for $N_c < N < N_e$, and finally (4) for 
sufficiently large energy, $N > N_e$, we should enter the regime of quantum 
ergodicity where (\ref{eq:qe}) holds.

\section{CANTORI AND QUANTUM MECHANICS}

Here we would like to illustrate explicitly the effect of cantori
on quantum eigenfunctions.
In the regime where $x = \epsilon^{3/2} k < 1$
it is natural to expect that cantori will influence
the localization process. Indeed, as shown above,  in such situation 
the cantori act as 
perfect barriers, and the quantum system looks as if 
classically integrable \cite{PRA}. It is therefore expected that the 
localization length of eigenstates must be of the order of the size 
of cantori.  On the other hand, since for small 
$\epsilon$ the cantori border can be much higher than the perturbative 
border, we may have here a nice possibility to study the effect of cantori
in quantum mechanics.  The average size of cantori has been found 
analytically and numerically to be (see section II, eq. (\ref{eq:sizeofc}))
\begin{equation}
\ell_c = c \epsilon l_{\max} = c \epsilon \sqrt{2E}
\label{eq:lc}
\end{equation}
with $c$ a numerical constant.
In figure 2b we show an eigenstate in angular momentum basis
in the cantori region near the 2:1 resonance.
In figure 3 (and also figure 4) our numerical data  show that at $x=1$ 
the behaviour of localization length changes: for $x<1$ namely below the cantori 
border the localization length agrees with the theoretical estimate (\ref{eq:lc}).
This provides the first numerical evidence that the
idea of flux quantization  through cantori introduced in\cite{MCKAY}
is indeed correct.
Even more convincing is the inspection of individual quantum
eigenstates in the Husimi phase space representation (see 
figure 5) which show very clearly, that in the cantori region, the
relative phase space area occupied by the localized eigenstates does not
increase with increasing $\epsilon$ or with increasing $k=\sqrt{2E}$. 
We checked this even quantitatively, by computing the phase space
area of Husimi phase space distributions via information
entropy or inverse participation ratio \cite{Prosen96}.
The width of Husimi functions on quantized cantori in the perpendicular 
direction is given only by the width of the wavepacket which is used 
for the computation of Husimi functions (see section VI).

\section{INTERPLAY BETWEEN CANTORI AND DYNAMICAL LOCALIZATION}

Above the quantum cantori border $x=1$ the flux through the turnstiles
 becomes larger than a Planck's
cell,  the cantori do not act any more as barriers for quantum dynamics 
 and the quantum motion starts to follow the classical diffusive 
behaviour up to the quantum relaxation time $t_R$ (break time), which 
is proportional to the density of {\em operative eigenstates} namely of those states
which enter the initial condition and therefore actually control the quantum dynamics.
For $t>t_R$ instead, the quantum dynamics enter an oscillatory regime around the 
stationary localized state. 
The density of operative eigenstates is by a factor 
$\ell/l_{\max}$ smaller than the total density of states
 $$t_R = (\ell/l_{\max}) dN/dE$$ 
and therefore the relaxation time is less than the Heisenberg time which is given
by the level density. The  angular momentum width $\ell$ of the localized state is
then given by $\ell^2 \approx D t_R/T$, where $T\approx E^{-1/2}$ is the average
time between  bounces. This leads to the simple expression for the
average (scaled) localization length
\begin{equation}
\ell/l_{\max} \approx D/k = \alpha \epsilon^{5/2} k
\label{eq:dynloc}
\end{equation}
where $\alpha$ is a constant to be determined numerically.
     We need however to take into account the fact that for $x<1$
the width of eigenfunctions is determined by classical cantori and 
not by dynamical localization. Since we measure the localization 
length in angular momentum space, then for $x>1$, we need to 
add the average size of cantori to the angular
momentum spread due to dynamical localization. Therefore, above the 
cantori border $x > 1$, we expect the 
following expression for the numerical (scaled) localization length
$\sigma_n = \ell_n/l_{\max}$,
\begin{equation}
\sigma_n = \bar{p} + (1-\bar{p})\alpha \epsilon (x-1)
\label{eq:resc}
\end{equation}
which takes into account the fact that we need to rescale the
total size of angular momentum space, and that for $x=1$, 
$\sigma_n=\bar{p}$ which is the average size of cantori as determined
numerically.
In figure 3 (see also figure 4) it is seen that expression
(\ref{eq:resc}) is in excellent agreement with our numerical data
up to $N$  $\sim 10^7$ and for different values of $\epsilon$.
The obtained value of the numerical constant is
$\alpha=1.7$.

Notice that only for very small $\epsilon$ 
 the constant contribution due to cantori in (\ref{eq:resc}) 
will be negligible and
localization length would be simply proportional to $D$, 
as given by eq. (\ref{eq:dynloc}).

Notice also that due to the discontinuity of the stadium map, the quantum
eigenfunctions are not exponentially localized, like for the standard map
or for smooth chaotic billiards 
\cite{FS1}, but they have  power-law tails 
$$p(l) \propto |l-l_0|^{-4}$$ 
This is consistent  with  results based on 
the quantization of the stadium map \cite{BORGO}, and with
rigorous results on  band random matrices with increasing
band size \cite{Mirlin}.
In figure 6 we show two (averaged) localized states in angular momentum 
basis in which the power law tails are clearly seen.

The pseudo-integrable (cantori) regime may also be detected by 
inspecting the energy level statistics, e.g. the 
commonly studied nearest neighbour level spacing distribution $P(S)$, 
the delta3 statistics etc. As an example
in figure 7 we show that $P(S)$ is nearly Poissonian in cantori regime
while it is intermediate between Poisson and Wigner (GOE) 
in the regime of dynamical localization.
We would like to stress that the above deviations from GOE predictions
have no relation with periodic orbit theory and bouncing ball orbits.

\section{LOCAL DENSITY OF STATES}

The Hamiltonian $H$ of the stadium billiard may be written as a small
deformation of the integrable circle billiard $H_0$, namely
$H = H_0 + \epsilon V$.
The exact stadium-eigenstates may be expanded in terms of
unperturbed quarter-cirlce eigenstates $\Phi_{s m}(\vec{r})$ 
where $k^0_{sm}$ are the eigenvalues of the integrable quarter-circle -- 
the zeros of the even-order Bessel functions (\ref{eq:express}).
Or, vice versa, we can express eigenstates of integrable quarter-circle
in terms of exact eigenstates of the perturbed billiard by the
same (orthogonal) matrix of coefficients $c^n_{sm}$,
$$\Phi_{sm} = \sum_{n} c^{n}_{sm} \Psi_{k_n}.$$

It is important to note that unlike for Wigner-band-random matrices
\cite{CCGI96} the matrix of coefficients $c^{n}_{sm}$,  ordered 
with increasing  wavenumber (energy), 
 has been found to have a symmetric appearance (see figure 8a
for an example). The structure of rows (expansions of exact states in
terms of circle states) is very similar to the structure of
columns (expansions of circle states in terms of exact states).
The effective bandwidth $b$ of the matrix $c^n_{sm}$ determines
the width of the energy shell $\delta E = b$ while the effective number
of nonzero entries in each row (fixed $n$) is proportional to the 
localization length $\ell$ of that state $\Psi_n$.
Such a symmetry between perturbed and unperturbed states seems to be
generic for conservative Hamiltonian systems. Indeed, a similar structure
of  the  matrix of coefficients has been found for the chaotic rough billiards 
(with shapes of wiggled circles) introduced in \cite{FS1} as well.
However, the bandwidth $b$ for the stadium billiard has been found to take 
always its maximal value, that is 
$$b \approx l_{\rm max} = \sqrt{2E},$$ 
independent of the parameter $\epsilon$, except in the
perturbative regime $N < N_p$ where it becomes smaller.
This should be contrasted with chaotic rough billiards 
\cite{FS1} for which, 
below the so called Wigner-ergodicity  border, the bandwidth $b$ 
has been found to decrease as $b\approx \epsilon^2 N$.

Below the ergodicity border, $N < N_e$, 
$\ell < l_{\max}$, the matrix 
$c^n_{ms}$ is sparse in both horizontal and vertical directions
(see figure 8a). (Note that the same is true also for the rough billiard) 
As a consequence of that, the averaged local density of states (LDOS),
$$w(k) = \ave{\sum_{n} |c^{n}_{sm}|^2 \delta(k + k^0_{sm} - k_n)}_{sm}$$
(averaged columns shifted to the same center) is nearly 
the same as average EF in circular basis
$$W(k) = \ave{\sum_{sm} |c^{n}_{sm}|^2 \delta(k + k_n -
\bra{n}\sqrt{2 H_0}\ket{n})}_{n}$$ 
(average rows shifted to the same mean). 
See figure 8b. Both distributions, $W(k)$ and $w(k)$, 
agree quite well with the theoretical distribution
$W_e (k) = (\sin(k)/k)^2/\pi$
which has been proposed\cite{FS1} for LDOS of nearly
circular billiards in the regime of quantum (Shnirelman)
ergodicity.
 
Such structure is found in the regime of dynamical localization
as well as in the regime of cantori localization where the
quantum system behaves as if  classically integrable.
We would like to stress that for the case of WBRM there 
is instead a strong asymmetry in the sense that the EF's are solid and narrow, namely they are localized inside the energy shell, while the expansion of a basis state in terms of exact EF's  shows a sparse structure \cite{CCGI96}.

\section{NUMERICAL METHODS}

\subsection{Computation of eigenstates of quantum stadium billiard}

In order to check the above theoretical predictions we had to 
compute numerically the quantum eigenfunctions $\Psi_n(\vec{r})
= \Psi_{k_n}$
of chaotic billiards which are solutions of the Schr\" odinger
equation $(\Delta + k^2_n)\Psi_n = 0$ (where $\hbar=1$) 
satisfying Dirichlet boundary conditions along the boundary.
We had to compute eigenstates with extremely large sequential
quantum numbers $N$ of the desymmetrized stadium billiard (and also
of the rough billiards) up to $N\approx 10^7$.

We have used a recently proposed scaling method \cite{VS95},
but we expanded quantum eigenfunctions $\Psi_k(\vec{r})$ in terms
of more suitable circular waves (here we consider only odd-odd states)
$$
\Psi_k(\vec{r}) = \sum_{s=1}^{M} a_s J_{2s}(k r)\sin(2s\theta)
$$
instead of the originally proposed plane waves.
The eigenvalue $k=\sqrt{2E}$ is determined by minimizing
a certain positive bilinear form along the boundary of the
billiard \cite{VS95}. This can be done by solving a generalized
eigenvalue problem of dimension $\approx k$ which yields
$\sim 10\% k$ accurate eigenvalues $k_n$ and eigenvectors $a^n_s$
simultaneously. Therefore, with this method, the computer workload required per energy 
level is by a factor  $k$ or even $10k$ (namely
from $10^3$ to $10^4$ for $k=10^3$) smaller than with more traditional 
methods, like the original Heller's plane wave decomposition or boundary 
integral method!

\subsection{Definition and computation of localization length}
 
Note that the magnitude of coefficients $a^n_s$ 
is proportional to the probability (\ref{eq:pn})
of having angular momentum
$l=2s$ in a quantum state $\Psi_n$,
$$
p_n(l=2s) = |a_s|^2 \int_0^1 dr r |J_{2s}(kr)|^2
\propto |a_s|^2\sqrt{1-l^2/k^2}.$$
The number of levels below a given  wavenumber $k$ or energy $E$, in a
desymmetrized billiard, is given by the Weyl formula $N = k^2/16 = E/8$.

The effective spread in angular momentum of an eigenstates $\Psi$
is characterized by the localization length 
$\ell$ which  measures the typical number of
$p_n(l)$ values which are substantially different from zero,
or the size of the angular momentum interval in which $p(l)$ is
significantly different than zero.
Of course, there is no a unique definition of localization length.
In particular the choice is quite delicate for the
quantum stadium where localization is algebraic 
unlike  billiards with smooth 
boundary where localization is expected to be exponential.
We have therefore considered four different working definitions:
(i) spread in angular momentum 
$$\ell_1 = \beta_1(\bra{\Psi}\op{l}^2\ket{\Psi} 
- \bra{\Psi}\op{l}\ket{\Psi}^2)^{1/2},$$
(ii)  information entropy
$$\ell_2 = \beta_2 \exp\left(-\sum_l p(l)\ln p(l)\right)$$
(iii) inverse participation ratio,
$$\ell_3 = \beta_3 (\sum_l [p(l)]^2)^{-1},$$
(iv) $99\%$ probability localization length
\begin{equation}
\ell_4 = \beta_4
\min\{\#{\cal A};\sum_{l\in{\cal A}} p_k(l) \ge 0.99\}.
\label{eq:ll}
\end{equation}
The numerical constants $\beta_j,j=1,2,3,4$ have been determined
with the condition that $\ell_j$ should take  the maximal value
$l_{\max}=k=\sqrt{2E}$ in the regime of quantum ergodicity.
All the four definitions of localization length give {\em qualitatively}
the same results.
However, the $99\%$ localization length (\ref{eq:ll})  
 which is proportional to the minimal number
of angular momentum eigenvalues that are needed to support
$99\%$ probability,
is the least sensitive to the slowly (algebraically) decaying tails
of the distribution $p(l)$ and has given the sharpest
and less fluctuating numerical results.
The results shown in figures $3$ and $4$ were obtained
using the definition (\ref{eq:ll}) of localization length with the
numerical value$\beta_4=1.38$.
 
\subsection{Computation of Husimi functions}

In figure 5 we show Husimi functions of (quarter) stadium eigenstates
on the Poincar\' e-Birhkoff section, with coordinates $(\varphi,l)$.
The desymmetrized phase space is a rectangle 
$(\varphi,l) \in [0,\pi/2]\times[0,k], k=\sqrt{2E}$.
We consider (almost) minimal wavepackets $\ket{l\theta}$ 
at phase space point $(\theta,l)$, which in the angle 
representation $\ket{\varphi}$, $0\le \varphi < 2\pi$ read
\begin{equation}
\braket{\varphi}{\theta,l} = 
C\exp\left(-\frac{2a}{k}\sin^2(\textstyle{\frac{1}{2}}
(\varphi-\theta))+il\varphi\right)
\end{equation}
where $C$ is a normalization constant and $a$ is the aspect ratio.
We take $a=1$ which means that the wavepacket has the same 
relative size in both directions.
The Husimi function $h(\varphi,l)$ of a billiard eigenstate 
$\Psi(\vec{r})$ is computed w.r.t. the normal derivative of an 
eigenfunction  along the boundary $\vec{r}(\varphi)$,
$\partial_n \Psi = (d\vec{r}/d\varphi)\wedge \nabla\Psi/||d\vec{r}/d\varphi||$,
\begin{equation}
h(\theta,l) = |\int d\phi \braket{\phi}{\theta,l}\partial_n 
\Psi(\vec{r}(\phi))|^2,
\label{eq:hus}
\end{equation}
where the numerical integration along the boundary of the billiard 
has been performed accurately using
high-order Gaussian quadratures.

\section{CONCLUSIONS}
The billiard in a stadium is a very clean mathematical model
of classical chaos and it is therefore particularly convenient to
study the modifications that quantum mechanics introduces
in our general picture of deterministic chaos. While the classical motion is
ergodic and mixing for any non zero values of the parameter $\epsilon$, the
quantum motion, as we have seen in this paper, exhibits a very rich structure
and different regimes of motion as a function of the parameter $\epsilon$ or
the energy $E$.
Two are the main points we
would like to stress:
a) classical cantori may have strong effects on quantum
mechanics. They lead to a new quantum border which is distinct from the
perturbative border and from the localization border. In the regime of 
{\em quantum cantori} the rescaled localization length $\ell/k$ does not 
depend on 
energy or wavenumber $k$. In other
words, the quantum dynamics is basically determined by the classical 
structure. This effect should be observable in real experiments.
b) The mechanism of localization is strictly connected to the sparsity of 
EF's when expanded on the basis of unperturbed circle states 
(or vice versa, sparsity  of unperturbed states when expanded in terms of 
the stadium EF's). When we
increase perturbation (or decrease $\hbar$), sparsity decreases, 
localization length of EF's increases until the delocalized or 
quantum ergodic regime is reached. 
We conjecture that this mechanism of localization is typical of conservative
Hamiltonian systems. Quite interestingly, the mechanism of localization
in WBRM is quite different and  is
associated to an asymmetry between EF's (solid and narrow) and 
unperturbed
states (delocalized  but sparse). 
In this sense,  WBRM cannot be taken to
represent the typical behaviour of classically 
chaotic conservative systems.

The main impulse to the present paper originated from a previous work of 
B.Chirikov\cite{LIN} in which a wiggled circle billiard was 
introduced as a model
for localization in conservative systems. Also we have benefit from  
several results
on the standard map. This confirms the importance and generality of the 
Chirikov standard map for the study of classical and quantum
deterministic chaotic motion. Since 30 years, generations of physicists
all around the world, have learned the beauty and complexity of nonlinear
dynamics on the base of this apparently simple map. We believe that
the Chirikov map is at the root of the impressive
growth of the whole field of nonlinear dynamics and chaos. Strangely enough
this fundamental contribution of B.V. Chirikov to one of the main 
achievements
of physics of this century has never been properly recognized. 
Several years
ago, in a friendly private letter to a colleague concerning a paper 
in which he
failed to give the necessary credit to Chirikov,  
a common friend Joseph Ford,
now deceased, wrote "..why not give to the old Russian bear his due?". 
We would like  to turn this question to the scientific community. 
  One of  the
authors of the present paper(GC),
would like to take the opportunity of this special issue of the journal
to express his  deep gratitude to his friend and teacher Boris 
Valerianovich Chirikov.

\vfill
\newpage

\section*{Figure caption}

\begin{figure}[htbp]
\hbox{\hspace{-0.1in}\vbox{
\hbox{
\leavevmode
\epsfxsize=3.5in
\epsfbox{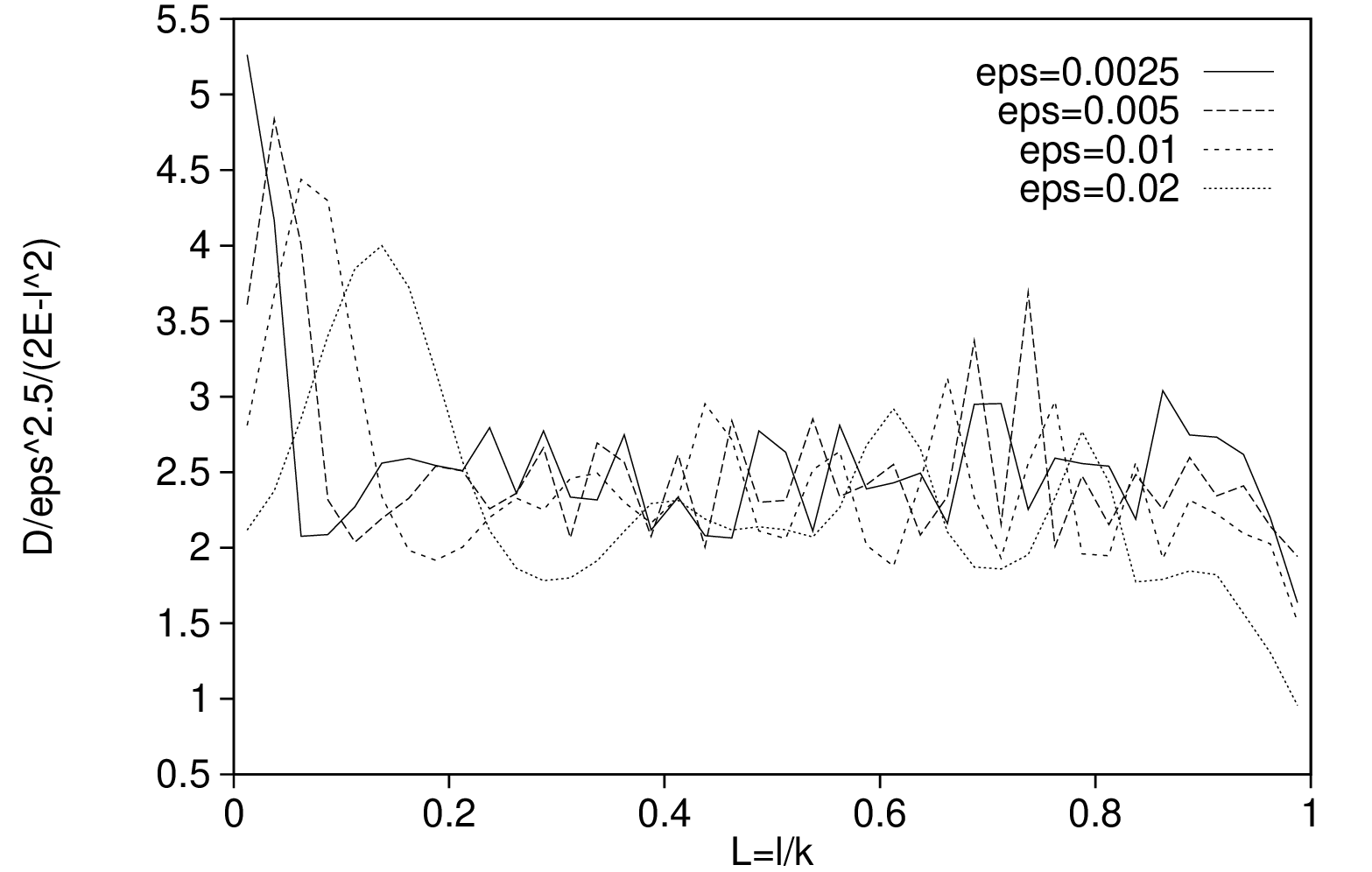}}
}}

\caption{
The rescaled classical diffusion rate (4)
$D(l)/[\epsilon^{5/2}(2E - l^2)]$ as a function of the
local value  $L=l/\sqrt{2E}$ of the angular momentum 
for different values of the control parameter
$\epsilon$.}

\label{fig:1}

\end{figure}

\begin{figure}[htbp]
\hbox{\hspace{-0.1in}\vbox{
\hbox{
\leavevmode
\epsfxsize=3.5in
\epsfbox{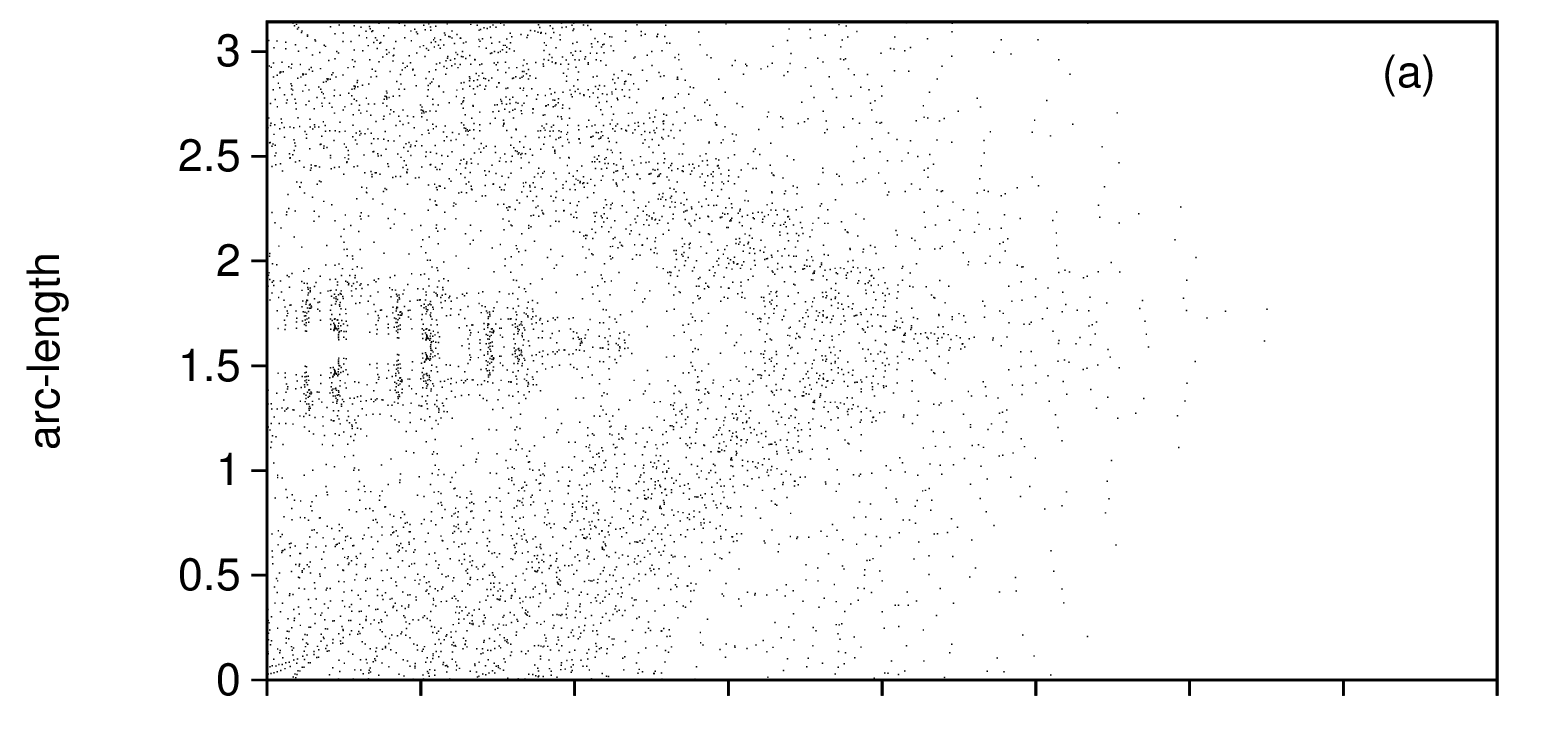}}
\vspace{-0.15in}
\hbox{
\leavevmode
\epsfxsize=3.5in
\epsfbox{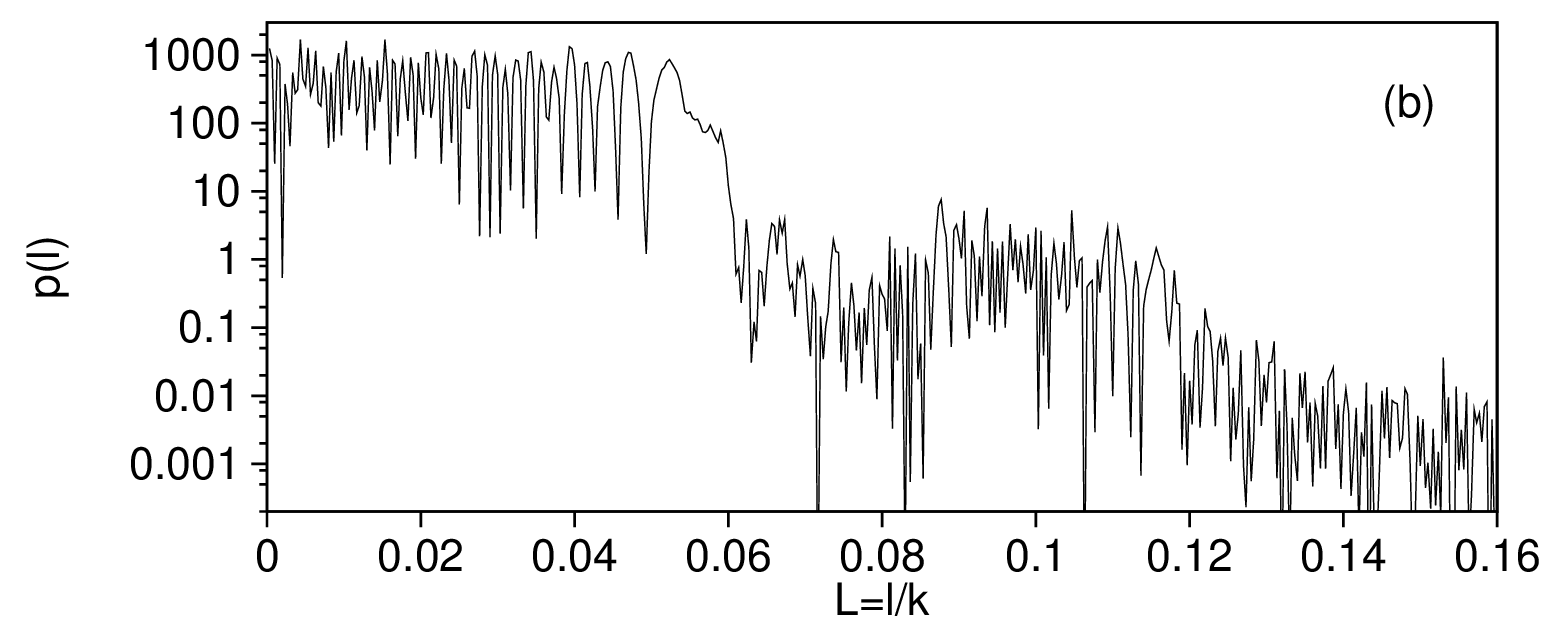}}
}}
\caption{
Time evolution of a single classical orbit, followed up to $20,000$
 bounces, for the classical billiard with $\epsilon=0.003$. The orbit is
initially started in the
middle of the largest `island' ( $L=0, \theta=\pi/4-0.0016$) (fig 2a).
Angular momentum probability distribution $p(l)$
of the corresponding  eigenstate 
with $\epsilon=0.003$ and eigenvalue $k=5999.8166$. 
As it is seen  the state is uniformly distributed over the cantorus
in the main island (Fig.2b). Notice also 
the same eigenstate in Husimi phase-space
representation (2nd state in Fig.5b).}

\label{fig:2}

\end{figure}

\begin{figure}[htbp]
\hbox{\hspace{-0.1in}\vbox{
\hbox{
\leavevmode
\epsfxsize=3.5in
\epsfbox{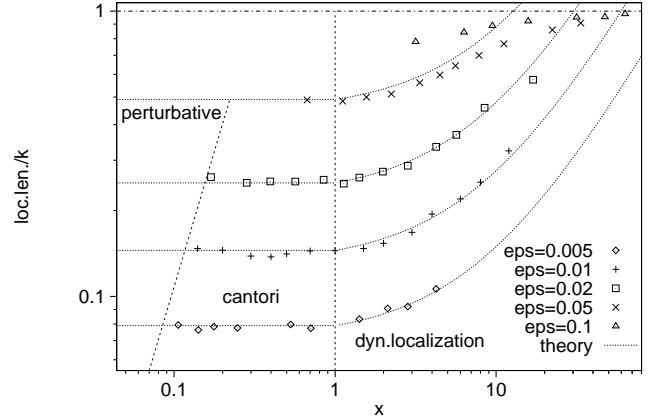}}
}}
\caption{Rescaled localization length $\sigma=\ell/k$
versus the scaling variable $x=\epsilon^{3/2} k$ for five different values of
$\epsilon$ ($60 < k < 12,000$).
Each point is obtained by averaging over a large number $\nu$ of 
consecutive eigenstates ($\nu = 100$ for small $k$ and $\nu = 1,000$ for 
large $k$).
The numerical data clearly show the cantori border $x=1$.
In the cantori region $\ell/k$ is constant as expected, while for $x>1$
the numerical data agree with the theoretical prediction
(\ref{eq:resc})(dotted curves). For large $x$, the value of localization
length $\sigma$ approaches the maximal ergodic value $\sigma=1$.}

\label{fig:3}

\end{figure}

\begin{figure}[htbp]
\hbox{\hspace{-0.1in}\vbox{
\hbox{
\leavevmode
\epsfxsize=3.5in
\epsfbox{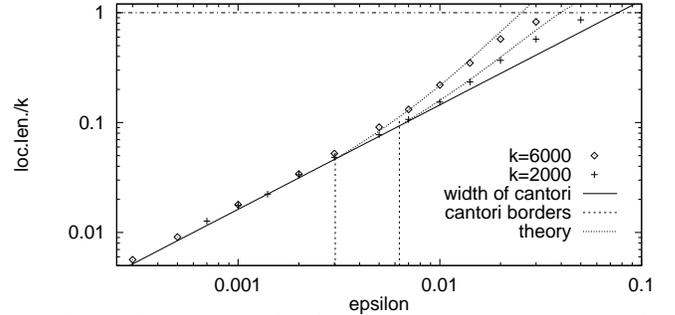}}
}}
\caption{Rescaled localization length $\sigma$ versus 
$\epsilon$ for two different wavenumbers $k$.
The full lines give the classical estimate for the average width of 
cantori.
It is seen that below the cantori border $x=1$($\epsilon=0.003$ for $k=6,000$
and $\epsilon=0.0063$ for $k=2,000$) the localization length is proportional to 
$\epsilon$ and independent on $k$. Above the border $x=1$ instead, the numerical data
follow the analytical estimate (15) (dotted curves).}

\label{fig:4}

\end{figure}

\vfill
\newpage

\begin{figure}[htbp]

\vspace{1truein}
{\noindent\em Color figure 5 (a-d) in (Windows/OS2 bitmap format) can be obtained 
by e-mail request on prosen@fiz.uni-lj.si}
\vspace{1truein}

\caption{The Husimi phase space representation of quantum
eigenstates of the stadium for different values of 
parameters $\epsilon$ and $k=\sqrt{2E}$. 
The Husimi function of a normal derivative of an eigenfunction
along the boundary of desymmetrized (quarter) stadium billiard (section VII) 
is plotted (x-axes: angle $0\le \varphi\le \pi/2$, y-axes: angular momentum
$0 < l < \sqrt{2E}$). We show four consecutive 
eigenstates in the following order: up-left, up-right, down-left, down-right for
(a) $\epsilon=0.001$, $k=5999.9775$, $5999.9812$, $5999.9815$, $5999.9818$, 
$(N\approx 2.25\cdot 10^6)$,
(b) $\epsilon=0.003$, $k=5999.8138$, $5999.8166$, $5999.8165$, $5999.8175$,
(c) $\epsilon=0.01$, $k=6000.0079$, $6000.0104$, $6000.0112$, $6000.0121$,
and
(d) $\epsilon=0.01$, $k=12000.0030$, $12000.0040$, $12000.0041$, $12000.0051$,
$(N\approx 9\cdot 10^6)$.
Note that the second state in (b) is the same as in figure 2b.
The color scale (to the right of each figure) is proportional to
the logarithm of Husimi function (white is large, black is small,
entire color scale extends over 8 orders of magnitude, a factor of
$10^8$. The cases (a,b) are in cantori region and one should note 
that the sizes of bright regions are roughly the same, while
(c,d) are in the region of dynamical localization.}

\label{fig:5}

\end{figure}

\begin{figure}[htbp]
\hbox{\hspace{-0.1in}\vbox{
\hbox{
\leavevmode
\epsfxsize=3.5in
\epsfbox{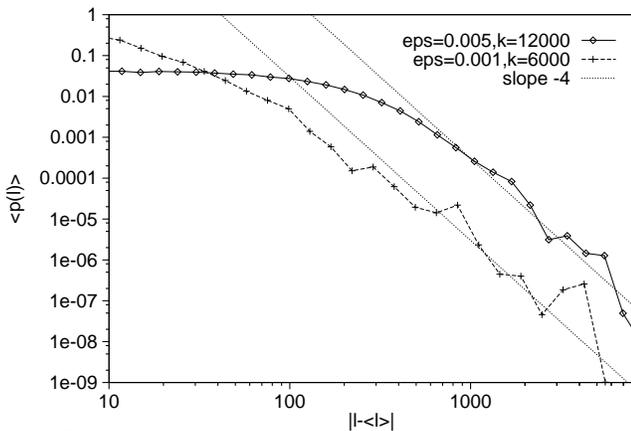}}
}}

\caption{ Average shape of eigenstates  in angular
momentum representation. (before averaging all the
eigenstates have been shifted to the same mean value $\ave{l}$ of angular
momentum. The average has been made over
about  thousand consecutive eigenstates for
$\epsilon=0.005,k\approx 12000$ (regime of dynamical localization),
and $\epsilon=0.001,k\approx 6000$ (cantori regime).
The dotted lines give the expected slope $-4$ for the tails.}

\label{fig:6}
\end{figure}

\begin{figure}[htbp]
\hbox{\hspace{-0.1in}\vbox{
\hbox{
\leavevmode
\epsfxsize=3.5in
\epsfbox{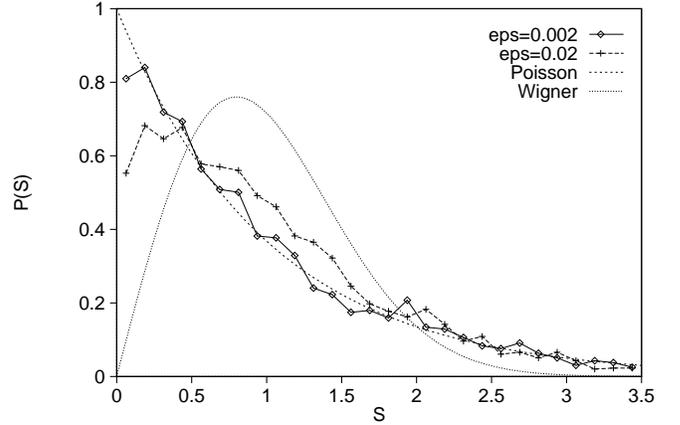}}
}}

\caption{ The nearest neighbour energy level spacing distribution $P(S)$.
We show data for the two spectral stretches in the same energy
window, $1993.8 \le k \le 2006.2$
($N \approx 250,000$),
containing about $3,100$ levels each, namely
$\epsilon=0.002$ (cantori region), and
$\epsilon=0.02$ (region of dynamical localization).
The theoretical Poisson and Wigner statistics are shown for comparison.
Notice that in the cantori region
$\epsilon=0.002$ (much above the perturbative border)
the numerical data closely follow  the Poissonian statistics
(the deviation is only slightly
larger than the expected statistical error).
}
\label{fig:7}
\end{figure}

\vfill

\newpage

\begin{figure}[htbp]
\hbox{\hspace{-0.1in}\vbox{
\hbox{
\leavevmode
\epsfxsize=3.5in
\epsfbox{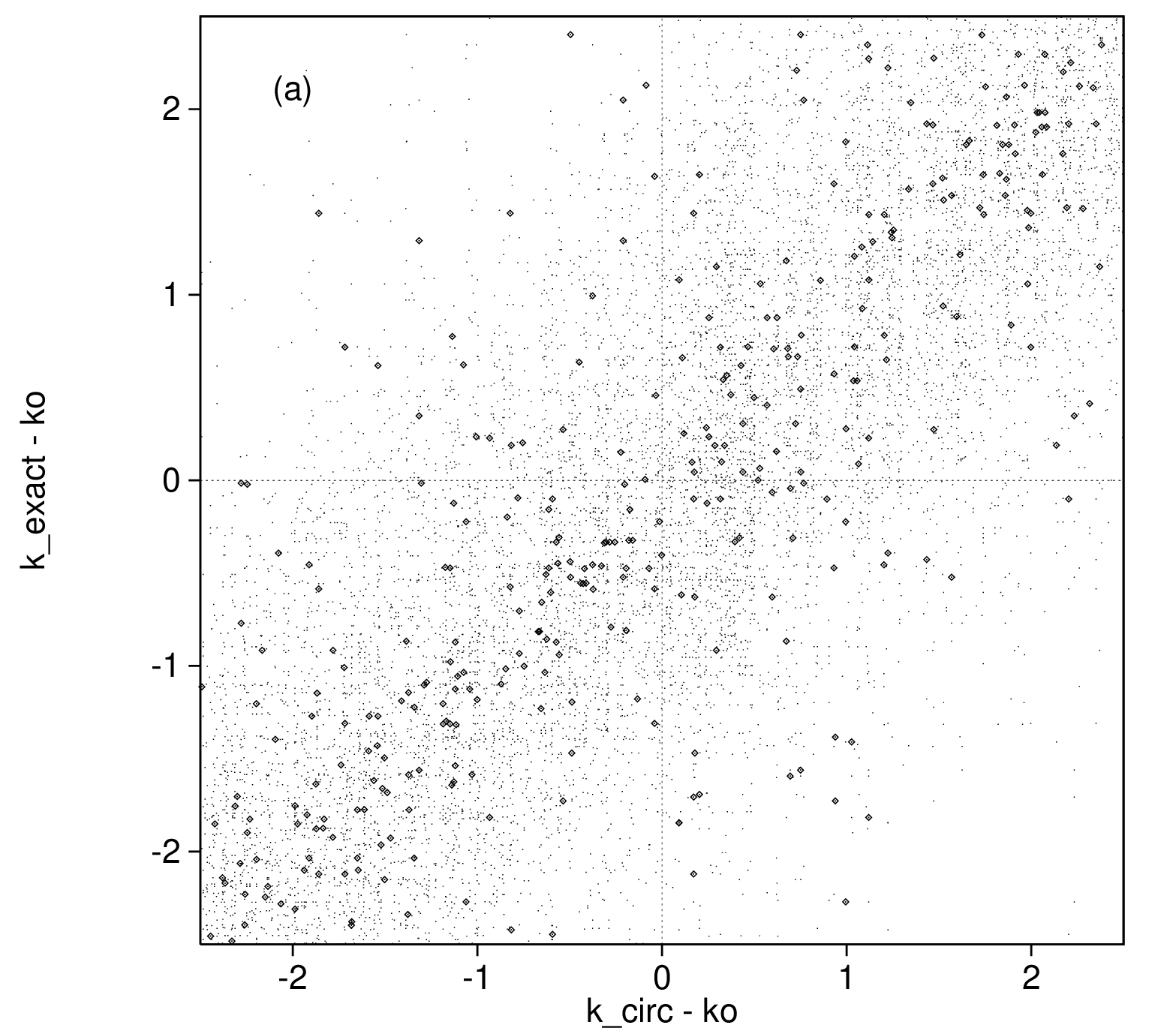}}
\hbox{
\leavevmode
\epsfxsize=3.5in
\epsfbox{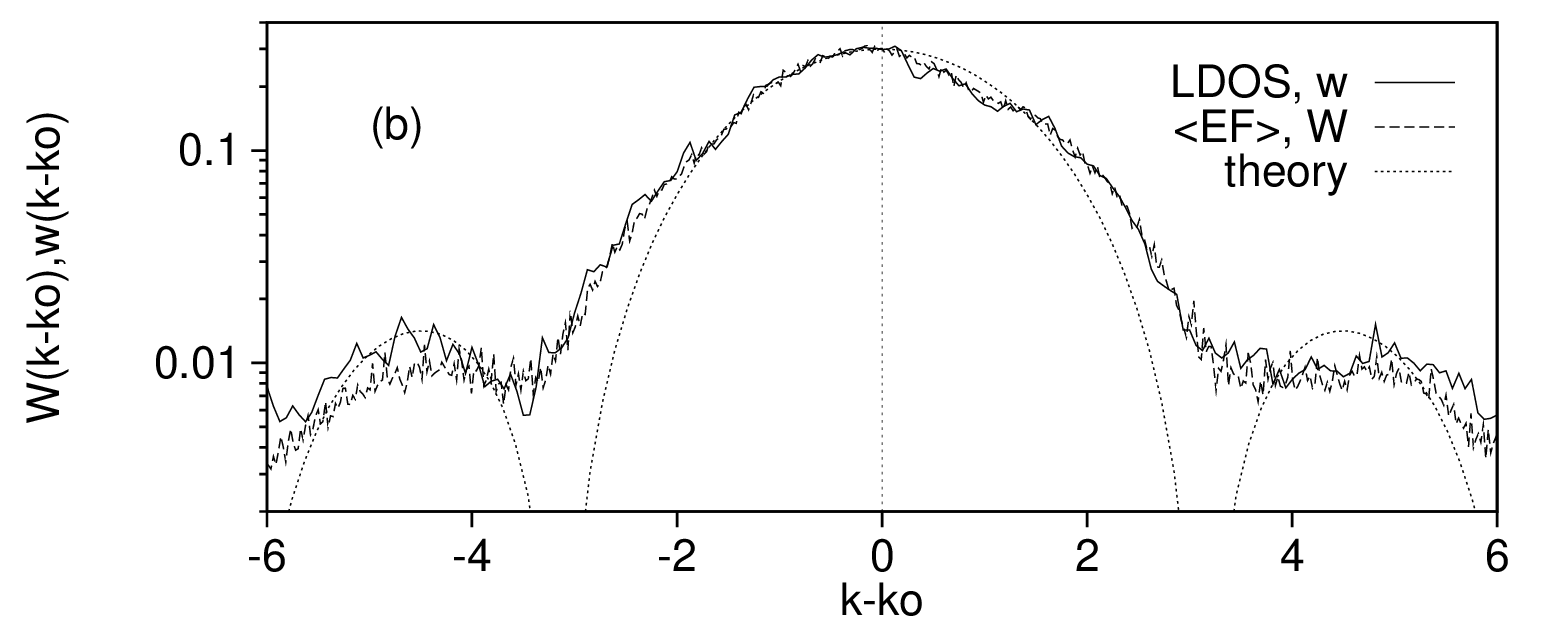}}
}}

\caption{ The structure of the matrix of coefficients $c^{n}_{sm}$
in the regime of dynamical localization:
$\epsilon=0.01$, $k\approx k_0 := 2000.0$ (Fig.8a)
A small dot at abscissa $k^0_{sm}-k_0$ and ordinate
$k_n-k_0$ is plotted if $|c^n_{sm}|^2 > 0.02$ and
a large dot if $|c^n_{sm}|^2 > 0.1$. Notice that the total number
of levels along each axis is about 1250.
In Fig.8b we plot the average local density of states $w(k-k_0)$ and
the average eigenfunction $W(k-k_0)$ for the same $\epsilon=0.01$,
averaged over a stretch of $5000$ consecutive eigenstates in the interval
$1990 < k < 2010$.
The dotted curve gives the theoretical LDOS for an ergodic
billiard $ W_e (k) = (\sin(k)/k)^2/\pi.$
}
\label{fig:8}
\end{figure}

\end{document}